\documentclass[prd,showpacs,twocolumn,preprintnumbers]{revtex4}

\usepackage{amsmath,amssymb,amsfonts,amsthm}
\usepackage{bm}
\usepackage{graphics}

\newcommand{\be}{\begin{equation}}
\newcommand{\ee}{\end{equation}}
\newcommand{\ba}{\begin{eqnarray}}
\newcommand{\ea}{\end{eqnarray}}
\def\bs{\begin{subequations}}
\def\es{\end{subequations}}
\def\a{\alpha}

\def\la{\lambda}
\def\k{\kappa}

\def\ve{\varepsilon}
\def\Om{\Omega}

\def\s{\sigma}

\def\vr{\varrho}

\def\cE{{\cal E}}

\def\cL{{\cal L}}

\def\cM{{\cal M}}

\def\p{\partial}
\def\B{\Box}
\newcommand{\Eq}[1]{(\ref{#1})}

\def\rme{e}
\def\rmd{d}
\def\rmi{i}

\begin{document}

\title{Fractal universe and quantum gravity}
\author{Gianluca Calcagni}
%\email{calcagni@aei.mpg.de}
\affiliation{Max Planck Institute for Gravitational Physics (Albert Einstein Institute)
Am M\"uhlenberg 1, D-14476 Golm, Germany}

\date{December 16, 2009}

\begin{abstract}
We propose a field theory which lives in fractal spacetime and is argued to be Lorentz invariant, power-counting renormalizable, ultraviolet finite, and causal. The system flows from an ultraviolet fixed point, where spacetime has Hausdorff dimension 2, to an infrared limit coinciding with a standard four-dimensional field theory. Classically, the fractal world where fields live exchanges energy momentum with the bulk with integer topological dimension. However, the total energy momentum is conserved. We consider the dynamics and the propagator of a scalar field. Implications for quantum gravity, cosmology, and the cosmological constant are discussed.
\end{abstract}

%\preprint{AEI-2009-125}
%\preprint{arXiv:0912.3142}
\preprint{Phys.\ Rev.\ Lett.\ {\bf 104}, 251301 (2010) \qquad [arXiv:0912.3142]}
\maketitle

%%%%%%%%%%%%%%%%%%%%%%%%%%%%%%%%%%%%%%%%%%%%%%%%%%%%%%%%%%%%%%%%%%%%%%%%%%%%%%%%%%%%%%%%%%%%%%%%%%%%%%%%%%%%%%%%%%%%%%%%%%%%%%%%%%%%%%%%%%%%%%%%%%%%%%%%%%%%%%%%%%%%%%%%%%%%%%%%%%%%%%%%%%%%%%%%%%%%%%%%%%%%%%%%%%%%%%%%%%%%%%%%%%%%%%%%%%%%%%%%

The search for a consistent theory of quantum gravity is one of the main issues in the present agenda of theoretical physics. In addition to major efforts such as string theory and (loop) quantum gravity, other independent lines of investigation have received some attention, including causal dynamical triangulations, asymptotically safe gravity, spin-foam models, and Ho\v{r}ava--Lifshitz (HL) gravity. All these theories exhibit a running of the spectral dimension $d_{\rm S}$ of spacetime such that at short scales $d_{\rm S}\sim 2$ \cite{spe}. Systems whose effective dimensionality changes with the scale can show fractal behaviour, even if they are defined on a smooth manifold.

By construction, HL gravity \cite{Hor2} surrenders Lorentz invariance as a fundamental symmetry. Because of the presence of relevant operators, the system is conjectured to flow from an ultraviolet (UV) fixed point to an infrared (IR) limit where, effectively, Lorentz and diffeomorphism invariance are restored at the classical level. However, loop corrections to the propagator of fields can lead to violations several orders of magnitude larger than the tree-level estimate, unless the bare parameters of the model are fine tuned  \cite{lor}. Despite the beautiful richness of its physics, the model is clearly under strong pressure, also for other independent reasons.

It is the purpose of this Letter to introduce an effective quantum field theory with two key differences with respect to HL gravity. The first is that power-counting renormalizability is obtained when the fractal behaviour is realized at the structural rather than effective level, i.e., when it is implemented in the very definition of the action. In other words, we will require not only the spectral dimension of spacetime, but also its UV Hausdorff dimension to be $d_{\rm H,UV}\sim2$. The second difference is that we wish to maintain Lorentz invariance. The proposal is mainly focused on a scalar field but we do not foresee any obstacle to extend it to other matter sectors or gravity. Most of the ingredients in our recipe are shared by other models, but their present mixing will hopefully give fresh insight into some aspects of quantum gravity. For example, a running cosmological constant naturally emerges from geometry as a consequence of a deformation of the Poincar\'e algebra. A considerably expanded presentation, bibliography, and proofs of the main statements and formul\ae\ are given in a companion paper \cite{fra2}.

\emph{Fractal universe.}---We require that time and space coordinates scale isotropically. In momentum units, $[x^\mu]=-1$ for $\mu=0,1,\dots,D-1$. The standard measure in the action is replaced by a nontrivial measure (which appears in Lebesgue--Stieltjes integrals \cite{CvB}),
\be\label{vr}
\rmd^D x\to \rmd\vr(x)\,,\qquad [\vr]=-D\a\neq -D\,.
\ee
Here $D$ is the topological (positive integer) dimension of the embedding spacetime and $\a>0$ is a parameter. What kind of measure can we choose? A two-dimensional small-scale structure is a desirable feature of renormalizable spacetime models of quantum gravity, and a na\"ive way to obtain it is to let the effective dimensionality of the universe change at different scales. We claim that a simple realization of this feature is the definition of a fractional action. Another route, which we shall not follow here, is to define particle physics directly on a fractal set with general Borel probability measure $\vr$. This was done in \cite{Svo87} (and \cite{Ey89} on Sierpinski carpets) for a quantum field theory on sets with $d_{\rm H}$ very close to 4.

The above claim is motivated by results in classical mechanics, according to which integrals on net fractals can be approximated by fractional integrals \cite{1dfr} which, in turn, are particular Lebesgue--Stieltjes integrals. The approximation is valid for large Laplace momenta and can be refined to better describe the full structure of the Borel measure $\vr$ characterizing the fractal set. The order $\a$ of a fractional integral describing a random process is related with the Hausdorff dimension of the process itself. Different values of $0<\a\leq 1$ mediate between full-memory ($\a=1$) and Markov processes ($\a=0$), and in fact $\a$ roughly corresponds to the fraction of states preserved at a given time during the evolution of the system. Applications of fractional integrals range from statistics to diffusing or dissipative processes with residual memory such as weather and stochastic financial models, to system modeling and control in engineering.

The classical mechanics results in one dimension can be easily generalized to a $D$-dimensional Euclidean space, thus opening a possibility for applications in spacetime. The most convenient form for defining a field theory Stieltjes action is a Weyl-type integral. Also in this case, the order of the fractional integral (for us, $D\a$) has a natural meaning as the Hausdorff dimension of the underlying fractal \cite{Tar1}. We shall see that for a Lorentzian integral the same interpretation holds, and that $D\a$ corresponds to the effective dimension of spacetime at a given scale. We do not have the theorems of the one-dimensional case at hand, but it is sufficient to limit ourselves to an operational definition of the Hausdorff dimension. Namely, $d_{\rm H}$ determines the scaling of a Euclidean volume (or mass distribution) of characteristic size $R$, $V(R) \sim R^{d_{\rm H}}$. Taking Eq.~\Eq{vr} on board,
\be
V(R)\sim\int_{D\textrm{-ball}}\rmd\vr_{\rm Eucl}(x)\sim \int_0^R \rmd r\, r^{D\a-1}\sim R^{D\a},
\ee
thus showing that 
\be
\a=\frac{d_{\rm H}}{D}.
\ee
An alternative definition of fractal dimension $d_{\rm F}$ entails the scaling properties of two-point correlation functions over an ensemble, $\langle\phi(x)\phi(y)\rangle\sim |x-y|^{2-d_{\rm F}}$. In our case $d_{\rm F}=d_{\rm H}=d_{\rm S}$, as we shall check for field Feynman propagators.

We formulate a scalar field theory with Stieltjes action for the purpose of controlling its properties in the ultraviolet. The model is interesting in its own right but also as a simple example whereon to work out the physics in preparation for the gravitational sector. In $D$ dimensions, we denote with $(\cM,\vr)$ the metric spacetime $\cM$ equipped with measure $\vr$ and consider the action
\be
S=\int_\cM\rmd\vr(x)\, \cL(\phi,\p_\mu\phi)\,,\label{stmes}
\ee
where $\cL$ is the Lagrangian density of the scalar field $\phi(x)$ and, if $\vr$ is absolutely continuous,
\be
\rmd \vr = (\rmd^D x)\,v(x) \label{stme}
\ee
is some multidimensional Lebesgue measure. Highly irregular sets with scale-dependent dimension do not satisfy the absolute continuity hypothesis, and they would be excluded \emph{a priori} if one had defined the model starting from Eq.~\Eq{stme} rather than \Eq{vr}. This is why a general formulation in terms of a Lebesgue--Stieltjes action is preferable over a Lebesgue action with weight $v$. Although sufficient for the goals of this paper, Eq.~\Eq{stme} might be too restrictive as soon as one wished to construct a concrete fractal support for the action. We assume $\cM$ to be a manifold but this may not be the case in general. Eqs.~\Eq{stmes} and \Eq{stme} resemble a field theory with a dilaton or conformal rescaling of the Minkowski determinant \cite{foot}.

Since we wish the Lorentz group $SO(D-1,1)$ to be part of the symmetry group of the action, the Lagrangian density $\cL$ and weight $v$ must be Lorentz invariant separately. The former can be taken to be the usual scalar field Lagrangian, $\cL=-\p_\mu\phi\p^\mu\phi/2-V(\phi)$, where $V$ is a potential and contraction of Lorentz indices is done via the Minkowski metric $\eta_{\mu\nu}= (-+\dots+)_{\mu\nu}$. As for the Stieltjes measure, we make a spacetime isotropic choice such that $[v]= D(1-\a)$.

We now pause and discuss the interpretation of the measure. Classically, one can boost solutions of the equation of motion to a Lorentz frame where $v=v({\bf x})$ (spacelike fractal) or $v=v(t)$ (timelike fractal). These two cases will lead to different physics but both corresponding to a dissipative system. This conclusion is in line with the known results of fractional mechanical systems and we shall make it explicit later. At the quantum level, all configurations should be taken into account, so there is no quantum analogue of space- or timelike fractals. The theory on the $D\a$-dimensional fractal is expected to be dissipative, i.e., nonunitary. Fortunately this will not be a problem because, from the point of view of the manifold with $D$ topological dimensions, energy is indeed conserved.

\emph{Renormalization.}---The scaling dimension of $\phi$ is $[\phi]=(D\a-2)/2$, which is zero if, and only if, $\a=2/D$. Then $d_{\rm H}=2$, and $\a=1/2$ in four dimensions. This value can change for other definitions of the measure \cite{fra2}. Let the scalar field potential be polynomial, $V=\sum_{n=0}^N \s_n\phi^n$. The coupling of the highest power has engineering dimension $[\s_N]=D\a-N(D\a-2)/2$. For the theory to be power-counting renormalizable $[\s_N]\geq 0$, implying $N\leq 2D\a/(D\a-2)$ if $\a>2/D$, and $N\leq+\infty$ if $\a\leq2/D$.
When $\a=1$, one gets the standard results $[\phi]=(D-2)/2$, $N\leq 2D/(D-2)$. These considerations induce us to try to have the parameters run from an ultraviolet nontrivial fixed point where $\a=2/D$ to an infrared fixed point where, effectively, $\a=\a_{\rm IR}$. The dimension of spacetime is well constrained to be 4 from particle physics to cosmological scales and starting at least from the last scattering era \cite{CO}. Therefore, $\a_{\rm IR}=1$ if $D=4$. To actually realize this particular renormalization group flow, one should add relevant operators to the action corresponding to terms with trivial measure weight. Then the total scalar action is $S=\int\rmd^D x[v\cL+M^{D(1-\a)}\tilde\cL]$, where $M$ is a constant mass term and $\tilde\cL$ is $\cL$ with all different bare couplings ($\s_n\to\tilde\s_n$). We symbolically represent this modification of the action as
\be\label{mea}
v(x)\to v(x)+M^{D(1-\a)}\,.
\ee
The constant term is anyway required in the most general Lorentz-invariant definition of the measure weight.

Of course, this construction falls short of demonstrating the existence and effectiveness of such a flow, which should be verified by explicit calculations. Our attitude will be to introduce the model and first see its characteristic features and possible advantages, leaving the issue of actual renormalizability for the future. Anyway, at the classical level the system does flow from a lower-dimensional fractal configuration to a smooth $D$-dimensional one. This is clear from the definition \Eq{mea} of the measure weight and its scaling properties when $\a<1$. At small space-time scales, the weight $v\sim |x|^{D(\a-1)}$ dominates over the constant term, while at large scales or late times it is negligible. This is true simply by construction, and independently from renormalization issues. Therefore, at least the phenomenological effectiveness of the model is guaranteed.

\emph{Dynamics.}---The Euler--Lagrange and Hamilton equations of classical mechanical systems with (absolutely continuous) Stieltjes measure have been discussed in \cite{El05} in the one-dimensional case and \cite{ElT} in many dimensions. We can easily adapt the same procedure. From now on we consider only the UV part of the action, setting $M=0$. Any result in the infrared can then be obtained by going to the effective limit $\a\to 1$.

Now that the metric space is equipped with a nontrivial Stieltjes measure, caution should be exercised when performing functional variations. For instance, the correct Dirac distribution is such that $1=\int \rmd\vr(x)\,\delta_v^{(D)}(x)$, as was also noticed in \cite{Svo87}. The principle of least action yields the equation of motion
\be
\B\phi+\frac{\p_\mu v}{v}\p^\mu\phi-V'=0\,,
\ee
where $\B=\p_\mu\p^\mu$ and a prime denotes differentiation with respect to $\phi$. The above friction term is characteristic of dissipative systems and one would expect the Noether current associated with the usual Lagrangian continuous symmetries not to be conserved. On the other hand, one can easily find generalized conserved currents. In fact, the continuity equation for the energy-momentum tensor $T^\mu_{\ \nu}\equiv -\p_\nu\phi\,\p\cL/\p(\p_\mu\phi)+\delta^\mu_\nu\cL$ is $\p_\mu(v T^\mu_{\ \nu})-\cL\p_\nu v=0$.
Integrating this equation in space,
\be
\dot P_{\nu}+\int\rmd{\bf x}\,\p_\nu v\,\cL =0\,,\label{noe2}
\ee
where $P_\nu\equiv-\int\rmd{\bf x}\, v\, T^0_{\ \nu}$. The $\nu=i=1,\dots,D-1$ components give the conservation law of the physical momentum. The $\nu=0$ component yields conservation of the quantity ($t=x^0$)
\be
\cE(t) = H(t)+\Lambda(t)=H(t)+\int^t\rmd t\int\rmd{\bf x}\,\dot v\cL\,,\label{Lam}
\ee
which we interpret as the energy of the system ($H$ is the usual Hamiltonian but with $v$ weight). $\Lambda$ acts as a running cosmological constant of purely geometric origin.

$P_\mu$ generates spacetime translations in the field but not in its conjugate momentum. The Poincar\'e algebra is now noncommutative unless $v$ is only time dependent:
\be\label{poia}
\{P_i,P_j\}_v=0\,,\qquad \{H,P_i\}_v=\int\rmd{\bf x}\,\p_i v\,\cL\,,
\ee
where we defined the equal-time Poisson brackets with nontrivial measure weight $v$. One can check that also the Lorentz algebra is deformed.

Eqs.~\Eq{noe2}--\Eq{poia} signal dissipation. Regardless of the choice of classical fractal, one would also have to face the issue of unitarity at the quantum level. Moreover, we need a physical interpretation of dissipation. It turns out that the latter helps to address the above concern.

Consider a $(D-1)$-dimensional box of size $l$ and spatial volume $l^{D-1}$. At the scale $l$, particles live effectively in $D\a$ spacetime dimensions. If $\a=1$, they occupy the whole phase space in the box. Otherwise, they must dissipate energy, since the energy of the configuration filling the entire topological volume is different from that of a configuration limited to the effective $D\a$-dimensional world. The total energy of the system $\cE$ in $D$ topological dimensions is conserved, but the energy $H$ measured by a $D\a$-dimensional observer is not.

The $D$-dimensional side of the picture can be actually made more precise. So far we have interpreted the function $v$ as (the derivative of) a Stieltjes measure defined on a fractal of Hausdorff dimension $D\a$. One can also regard it as a ``dilaton'' field coupled with the Lagrangian density $\cL$ living on a $D$-dimensional manifold. Then it is natural to consider the usual $\delta$ of Dirac and Poisson brackets $\{\,\cdot,\cdot\,\}_1$. In that case, one can see that the new conjugate momentum is $\pi_\phi=v\dot\phi$ and all explicit $v$ dependence disappears in the $D$-momentum: $P_\mu=P_\mu(\phi,\pi_\phi)$. Poincar\'e and Lorentz invariance are then preserved and, at least at the classical level, dissipation occurs relatively between parts of a conservative system. Quantization would follow through, although an UV observer would experience an effective probability flow through his world-fractal.

To summarize, from the point of view of the observer living in the fractal and measuring geometry with weight $v$, translation and Lorentz invariance are broken inasmuch as the Poincar\'e algebra is deformed. In other words, when talking about fractals embedded in Minkowski spacetime we mean \emph{the geometries defined by the deformed Poincar\'e group}. However, from the point of view of the ambient $D$-dimensional manifold Poisson brackets and functional variations no longer feature the nontrivial measure weight, which is now regarded as an independent matter field. In that case, the full Poincar\'e group is preserved.

\emph{Propagator.}---The theory is Lorentz invariant and, in the ultraviolet, two-dimensional, so it is ghost free and causal at all scales. The Green function $G$ must depend only on the Lorentz interval $s^2=x_\mu x^\mu$. A calculation of the free partition function shows that $Z_0[J]=\exp\left[(i/2)\int \rmd\varrho(x)\rmd\varrho(y) J(x)G(x-y)J(y)\right]$, where $J$ is a source. The separate pointwise dependence on $x$ and $y$, as expected from the breaking of translation invariance, is absorbed in the measures rather than in $G$, which is a function only of the interval.

Since the causal propagator $G$ can be argued to be proportional, in configuration space, to the Green function of another well-known problem (the functional inverse of a fractional power of the d'Alembertian \cite{BGG}), we can already guess its pole structure in momentum space: in general, it will exhibit a branch cut with branch point at $k^2=-m^2$. Then, the theory is ghost free and its spectrum is a continuum of modes with rest mass $\geq m$. These expectations are fulfilled in an explicit calculation \cite{fra2}. In what follows, the substitutions $s^2\to s^2+\rmi\ve$ and $k^2\to k^\mu k_\mu-\rmi\ve$ are understood.

The massive propagator is of the form $G(s)\propto s^{D\a/2-1} K_{D\a/2-1}(ms)$, where $K$ is a modified Bessel function. When $\a=1$, $G$ is the usual Klein--Gordon propagator in $D$ dimensions. The propagator for timelike intervals is just the analytic continuation of the former. In the massless limit, for $s>0$
\be
G(s)=\frac{\rmi(s^2)^{1-\frac{D\a}2}}{\Om_D(2-D\a)}\ \stackrel{\a\to \tfrac2D}{\longrightarrow} G_*(s)=\frac{\Om_2}{\Om_D}\frac{\rmi}{\Om_2}\ln s\,,
\ee
where $\Om_D$ is the surface of the $(D-1)$-sphere. Both the propagator with $\a\neq 2/D$ and $\a=2/D$ display the promised scaling properties in dimension $d_{\rm H}=D\a$.

To calculate the propagator in momentum space, one has to take the Fourier--Stieltjes transform $\tilde G(k)$ of $G(s)$ in the above cases, defined with respect to $\vr$:
\be
G(x-y) = \frac{1}{(2\pi)^D}\int\rmd\vr(k)\,\rme^{\rmi k\cdot (x-y)}\tilde G(k)\,.\label{propk}
\ee
When $\a\neq 2/D$ and $m=0$, $\tilde G(k)=-[(D-2)/(D\a-2)]/k^2$. For general $\a>2/D$, the sign of the residue is always negative, which ensures the absence of ghosts. However, its value is not 1 ($\a=1$) but given by a geometric factor. This is expected as the effective theory in the world-fractal is not unitary and some probability is exchanged with the $D$-dimensional topological bulk. The massive and $\a=2/D$ cases are discussed in \cite{fra2}.

\emph{Gravity.}---The properties of the scalar field on an effective fractal spacetime are shared also by the gravitational sector. The Ansatz for its action is
\be\label{Sg}
S_g=\frac{1}{2\k^2}\int\rmd\vr(x)\,\sqrt{-g}\,\left(R-2\la\right)\,, 
\ee
where $g$ is the determinant of the dimensionless metric $g_{\mu\nu}$, $\k^2=8\pi G$ is Newton's constant, and $\la$ is a bare cosmological constant. The couplings have dimension $[\k^2]= 2-D\a$ and $[\la]=2$. In spacetime with $D=2$ topological dimensions and trivial measure weight $v=1$, the Einstein--Hilbert action is a topological invariant and there are no dynamical degrees of freedom. This is not the case for the theory \Eq{Sg} in the UV.

To describe the flow from the UV to the IR fixed point, we should add relevant operators also into the gravitational action. (The relevant operators in the matter sector are minimally coupled with gravity and they would not be enough.) This is done in the same way as for the matter sector, Eq.~\Eq{mea}, with possibly different $M_g\neq M_\phi$. The effective Newton constant then runs from a UV bare dimensionless constant to an IR value $\k^2_{\rm IR}\sim  \k^2_{\rm UV} M_g^{2-D}$. Note that $\k^2_{\rm IR}$ is not necessarily the observed Newton constant $\k^2_{\rm obs}$, in which case $M_g\sim m_{\rm Pl}$. As one can see from the equations of motion, $\k^2_{\rm obs}$ will depend on the background as well as on the scale of the problem.

\emph{Future developments.}---Several quantum field theory and cosmological implications of the model will require further study. If the system quickly flows to the IR fixed point, any direct effect from spacetime dimensionality might be negligible on cosmological spacetime scales, while it may be relevant in the early universe during inflation. Notably, at late times an imprint of the nontrivial short-scale geometry might survive as a cosmological constant, Eq.~\Eq{Lam}--- this should be of purely gravitational origin, since pressureless matter does not contribute.

%%%%%%%%%%%%%%%%%%%%%%%%%%%%%%%%%%%%%%%%%%%%%%%%%%%%%%%%%%%%%%%%%%%%%%%%%%%%%%%%%%%%%%%%%%%%%%%%%%%%%%%%%%%%%%%%%%%%%%%%%%%%%%%%%%%%%%%%%%%%%%%%%%%%%%%%%%%%%%%%%%%%%%%%%%%%%%%%%%%%%%%%%%%%%%%%%%%%%%%%%%%%%%%%%%%%%%%%%%%%%%%%%%%%%%%%%%%%%%%%


\begin{thebibliography}{99}

\bibitem{spe}   J. Ambj{\o}rn, J. Jurkiewicz, and R. Loll, Phys.\ Rev.\ Lett. {\bf 95}, 171301 (2005); O. Lauscher and M. Reuter, J.\ High Energy Phys.\ 10 (2005) 050; L.~Modesto, Classical Quantum Gravity {\bf 26}, 242002 (2009); D.~Benedetti, Phys.\ Rev.\ Lett. {\bf 102}, 111303 (2009); P.~Ho\v{r}ava, Phys.\ Rev.\ Lett.\ {\bf 102}, 161301 (2009).
\bibitem{Hor2}  P.~Ho\v{r}ava, Phys.\ Rev.\ D {\bf 79}, 084008 (2009).
\bibitem{lor}   J.~Collins, A.~Perez, D.~Sudarsky, L.~Urrutia, and H.~Vucetich, Phys.\ Rev.\ Lett.\ {\bf 93}, 191301 (2004); R.~Iengo, J.G.~Russo, and M.~Serone, J.\ High Energy Phys.\ 11 (2009) 020.
\bibitem{fra2}  G.~Calcagni, J.\ High Energy Phys.\ 03 (2010) 120.
\bibitem{CvB}   M. Carter and B. van Brunt, \emph{The Lebesgue--Stieltjes Integral: A Practical Introduction} (Springer, New York, 2000).
\bibitem{Svo87} K.~Svozil, J.\ Phys.\ A {\bf 20}, 3861 (1987).
\bibitem{Ey89}  G.~Eyink, Commun.\ Math.\ Phys.\ {\bf 125}, 613 (1989); G.~Eyink, Commun.\ Math.\ Phys.\ {\bf 126}, 85 (1989).
\bibitem{1dfr}  F.B.~Tatom, Fractals {\bf 3}, 217 (1995); F.-Y.~Ren, J.-R.~Liang, X.-T.~Wang, and W.-Y.~Qiu, Chaos Solitons Fractals {\bf 16}, 107 (2003).
\bibitem{Tar1}  V.E.~Tarasov, Chaos {\bf 14}, 123 (2004). 
\bibitem{foot} The present model is not just an exotic reformulation of dilaton scenarios for at least three major reasons. First, the dilaton of string theory couples differently in different sectors, while the scalar field $v$ appears as a global rescaling. Second, a change in the measure is consistently accompanied by a new definition of functional variations and Dirac distribution, in turn leading to an unfamiliar propagator and the deformation of the Poincar\'e algebra. Third, because of the foreign physical motivation $v$ must scale in a certain way in the UV. As a consequence, the physics is different in any respect \cite{fra2}.
\bibitem{CO}    F.~Caruso and V.~Oguri, Astrophys.\ J.\ {\bf 694}, 151 (2009).
\bibitem{El05}  R.A.~El-Nabulsi, Fizika A {\bf 14}, 289 (2005); G.S.F.~Frederico and D.F.M.~Torres, Int.\ J.\ Appl.\ Math.\ {\bf 19}, 97 (2006).
\bibitem{ElT}   R.A.~El-Nabulsi and D.F.M.~Torres, J.\ Math.\ Phys.\ (N.Y.) {\bf 49}, 053521 (2008); C.~Udri\c{s}te and D.~Opris, WSEAS Trans.\ Math.\ {\bf 7}, 19 (2008); R.A.~El-Nabulsi, Chaos Solitons Fractals {\bf 42}, 52 (2009); R.A.~El-Nabulsi, Chaos Solitons Fractals {\bf 42}, 2614 (2009).
\bibitem{BGG}   D.G.~Barci, C.G.~Bollini, L.E.~Oxman, and M.C.~Rocca, Int.\ J.\ Theor.\ Phys.\ {\bf 37}, 3015 (1998).
\end{thebibliography}
\end{document}